\documentclass[12pt]{elsart}

\usepackage{amsmath}
\usepackage{amsfonts}
\usepackage{graphicx}
\usepackage{amssymb}
\usepackage{ulem}

\begin{document}

\begin{frontmatter}

\title{Statistical magnitudes and the Klein tunneling \\ in bi-layer graphene: \\ 
influence of evanescent waves}

\author[jsr]{Jaime Sa\~{n}udo}
\ead{jsr@unex.es} and
\author[rlr]{Ricardo L\'{o}pez-Ruiz}
\ead{rilopez@unizar.es}

\address[jsr]{
Departamento de F\'isica, Facultad de Ciencias, \\
Universidad de Extremadura, E-06071 Badajoz, Spain, \\
and BIFI, Universidad de Zaragoza, E-50009 Zaragoza, Spain}

\address[rlr]{
DIIS and BIFI, Facultad de Ciencias, \\
Universidad de Zaragoza, E-50009 Zaragoza, Spain}


\begin{abstract}
The problem of the Klein tunneling across a potential barrier 
in bi-layer graphene is addressed. The electron wave functions are treated as 
massive chiral particles. This treatment allows us to compute the
statistical complexity and Fisher-Shannon information for each angle of incidence.
The comparison of these magnitudes
with the transmission coefficient through the barrier is performed. 
The role played by the evanescent waves on these magnitudes is disclosed.
Due to the influence of these waves, it is found that the 
statistical measures take their minimum values not only in the situations
of total transparency through the barrier, a phenomenon highly anisotropic for
the Klein \mbox{tunneling} in bi-layer graphene.  
\end{abstract}


\end{frontmatter}

\maketitle

In this work, we are concerned with the calculation of entropic magnitudes 
in a particular quantum scattering process through a potential barrier,
specifically the Klein tunneling in bi-layer graphene, a subject that 
can have interest in applied condensed-matter \cite{katsnelson2012}.

The crossing of potential barriers by quantum particles is a typical situation 
found in different quantum phenomena \cite{cohen}, such as tunneling \cite{hartman1962}, 
interferences \cite{perez2001}, resonances \cite{susan1994}, electron transport \cite{zhou2010}, etc.

A finite square barrier can be a first example where to study the relationship of these entropy-information 
measures with the reflection coefficient. In the context of non-relativistic quantum mechanics (NRQM), 
it has been put in evidence \cite{lopezruiz2013} that these statistical magnitudes present their minimum 
values just in the situations where the total transmission through the barrier is achieved.
A similar result was obtained for the crossing of a barrier in a relativistic-like quantum mechanics 
context \cite{sanudo2014}. 
This is the case of the so-called Klein tunneling \cite{klein1929} in single-layer graphene.

The Klein tunneling (or paradox)  is an exotic phenomenon with counterintuitive consequences in 
relativistic quantum mechanics (RQM) that has deserved a great 
interest in particle, nuclear and astro-physics \cite{sakurai1967,greiner1985,su1993,dombey1999,krekora2004,robinson2012}. 
Klein \cite{klein1929} showed that, in the frame of RQM, it is possible
that all incident particles (electrons in the case of graphene) can cross a barrier, independently of how high and 
wide the potential barrier is. This is a paradoxical fact from the point of view of the NRQM, 
where we know that the higher and wider the potential barrier is, less number of particles
can tunnel the barrier (exponential decay). 
The Klein paradox has never been realized in laboratory experiments. However, a possibility to observe this 
type of phenomenon has recently been proposed and tested by means of graphene \cite{geim2006,stander2009,young2009,he2013}.

From an electronic point of view, 
bi-layer graphene is a 2D gapless semiconductor with chiral electrons and holes with a finite mass $m$ \cite{geim2006}.
The mass of these quasiparticles is a non-relativistic consequence of their parabolic energy spectrum. The chirality comes 
from the spinorial nature of their wavefunction, similar to the chirality of the carriers in single-layer graphene that can 
be viewed as relativistic fermions, in that case massless due to their linear energy dispersion at low Fermi energies. 
For the present case of the bilayer graphene, the quadratic energy dispersion implies four possible solutions for a given energy $E$,  
\begin{equation}
E=\pm {\hbar^2 k_F^2 \over 2m},
\label{eqE}
\end{equation}
where $\hbar$ is the Planck's constant divided by $2\pi$, $k_F$ is the Fermi wavevector of the quasiparticle,
and $m\sim 0.035\,m_e$ is the effective mass of the carriers \cite{bai2007}, with $m_e$ the electron mass.
Two solutions correspond to propagating waves and the other two to exponentially growing and decaying waves
(evanescent waves). The expression of these wavefunctions can be found
by solving the off-diagonal Hamiltonian that describes the low-energy quasiparticles in bilayer graphene \cite{geim2006,macann2006}:

\begin{equation}
H_0=-{\hbar^2 \over 2m}
\left(\begin{array}{cc}
0 & (\hat{k}_x-i\hat{k}_y)^2 \\
(\hat{k}_x+i\hat{k}_y)^2 & 0
\end{array}\right),
\end{equation}

where $(\hat{k}_x, \hat{k}_y)=-i(\nabla_x,\nabla_y)$.

In order to study the Klein tunneling in bi-layer graphene, 
a two-dimensional square potential barrier $V(x)$ is considered
on the $x-y$ plane:

\begin{equation}
V(x) = \left\{
\begin{array}{rcl}
0, & \mbox{   }x\leq 0 & \mbox{(Region I),} \\
V_0, & \mbox{   }0<x<L & \mbox{(Region II),} \\
0, & \mbox{   }x\geq L  & \mbox{(Region III),} 
\end{array} \right.
\label{eq1}
\end{equation}

where $V_0$ and $L$ are the height and width of the barrier, respectively,
and the dimension along the $Y$ axis is supposed to be infinite. 
The effect of this potential barrier has the Klein tunneling as the most 
intriguing effect for $V_0>E$, where electrons outside the barrier transform 
into holes inside it, or vice versa, a phenomenon that allows to the charge carriers
to pass through the barrier. In the case of bi-layer graphene, this charge conjugation
requires the appearance of evanescent waves (holes with wavevector $ik$) inside the barrier.
This local potential barrier can be created by the electric field effect using local 
chemical doping or by means of a thin insulator \cite{novoselov2004,novoselov2005}.

Now, we consider an electron wave that propagates under the action of the Hamiltonian $H=H_0+V(x)$
at an incident angle $\phi$ respect to the $x$ axis. The solutions for the two-component spinor 
representing these quasiparticles in the different Regions are:

\begin{eqnarray}
\Psi_I(x,y) & = &  \left\{a_1\left(\begin{array}{c} 1 \\ se^{2i\phi}\end{array}\right)e^{ik_xx}
+ b_1\left(\begin{array}{c} 1 \\ se^{-2i\phi}\end{array}\right)e^{-ik_xx}\right. \nonumber \\
 & & \hspace{3.3cm}\left. + c_1\left(\begin{array}{c} 1 \\ -s\cdot h\end{array}\right)e^{\kappa_xx}\right\}\,e^{ik_yy},  \\
\Psi_{II}(x,y) & = &  \left\{a_2\left(\begin{array}{c} 1 \\ s'e^{2i\phi'}\end{array}\right)e^{ik'_xx}
+ b_2\left(\begin{array}{c} 1 \\ s'e^{-2i\phi'}\end{array}\right)e^{-ik'_xx}\right. \nonumber \\
 & & \hspace{0.5cm}\left. + c_2\left(\begin{array}{c} 1 \\ -s'\cdot h'\end{array}\right)e^{\kappa'_xx}
+ d_2\left(\begin{array}{c} 1 \\ -s'/h'\end{array}\right)e^{-\kappa'_xx} \right\}\,e^{ik_yy},  \\
\Psi_{III}(x,y) & = &  \left\{a_3\left(\begin{array}{c} 1 \\ se^{2i\phi}\end{array}\right)e^{ik_xx}
+ d_3\left(\begin{array}{c} 1 \\ -s/h\end{array}\right)e^{-\kappa_xx}\right\}\,e^{ik_yy},
\end{eqnarray}

where $k_x=k_{\mbox{\tiny F}}\cos\phi$,  $k_y=k_{\mbox{\tiny F}}\sin\phi$ are the wavevector components of the
propagative waves, $\kappa_x=k_{\mbox{\tiny F}}\sqrt{1+\sin^2\phi}$ is the decay rate of the evanescent wave, 
$k_{\mbox{\tiny F}}=\sqrt{2m|E|}/\hbar$ is the Fermi wavevector as given in expression (\ref{eqE}),
$s=sgn(-E)$, $h=(\sqrt{1+\sin^2\phi}-\sin\phi)^2$, outside the barrier. 
The values of the correspondent parameters inside the barrier are: 
$k'_x=k'_{\mbox{\tiny F}}\cos\phi'$,  $k'_y=k'_{\mbox{\tiny F}}\sin\phi'$, 
$\kappa'_x=k'_{\mbox{\tiny F}}\sqrt{1+\sin^2\phi'}$, $k'_{\mbox{\tiny F}}=\sqrt{2m|E-V_0|}/\hbar$,
$s'=sgn(V_0-E)$, $h'=(\sqrt{1+\sin^2\phi'}-\sin\phi')^2$, $\phi'=\arcsin{(\sqrt{|E|/|E-V_0|}\,\sin\phi)}$. 

The nine amplitudes ($a_1, b_1, c_1, a_2, b_2, c_2, d_2, a_3, d_3$) are complex 
numbers determined, up to a global phase factor, by the normalization condition and the boundary 
constraints, namely the continuity of both components of the spinor and their derivatives, at $x=0$ and $x=L$.
All of them can be numerically calculated. Remark that divergent exponential waves only take place in Region $II$ 
and the evanescent (exponentially decaying) waves appear in the three Regions.   

The scattering region (Region II) provokes a partial reflection of the incident electron wave. The reflection
coefficient $R$ gives account of the proportion of the incoming electron flux that is reflected by the barrier.
The expression for $R$ is:
\begin{equation}
R = {Flux_{reflected} \over Flux_{incident}}={|b_1|^2 \over |a_1|^2}\,.
\end{equation}

\begin{figure}[h]
\centerline{\includegraphics[width=7cm]{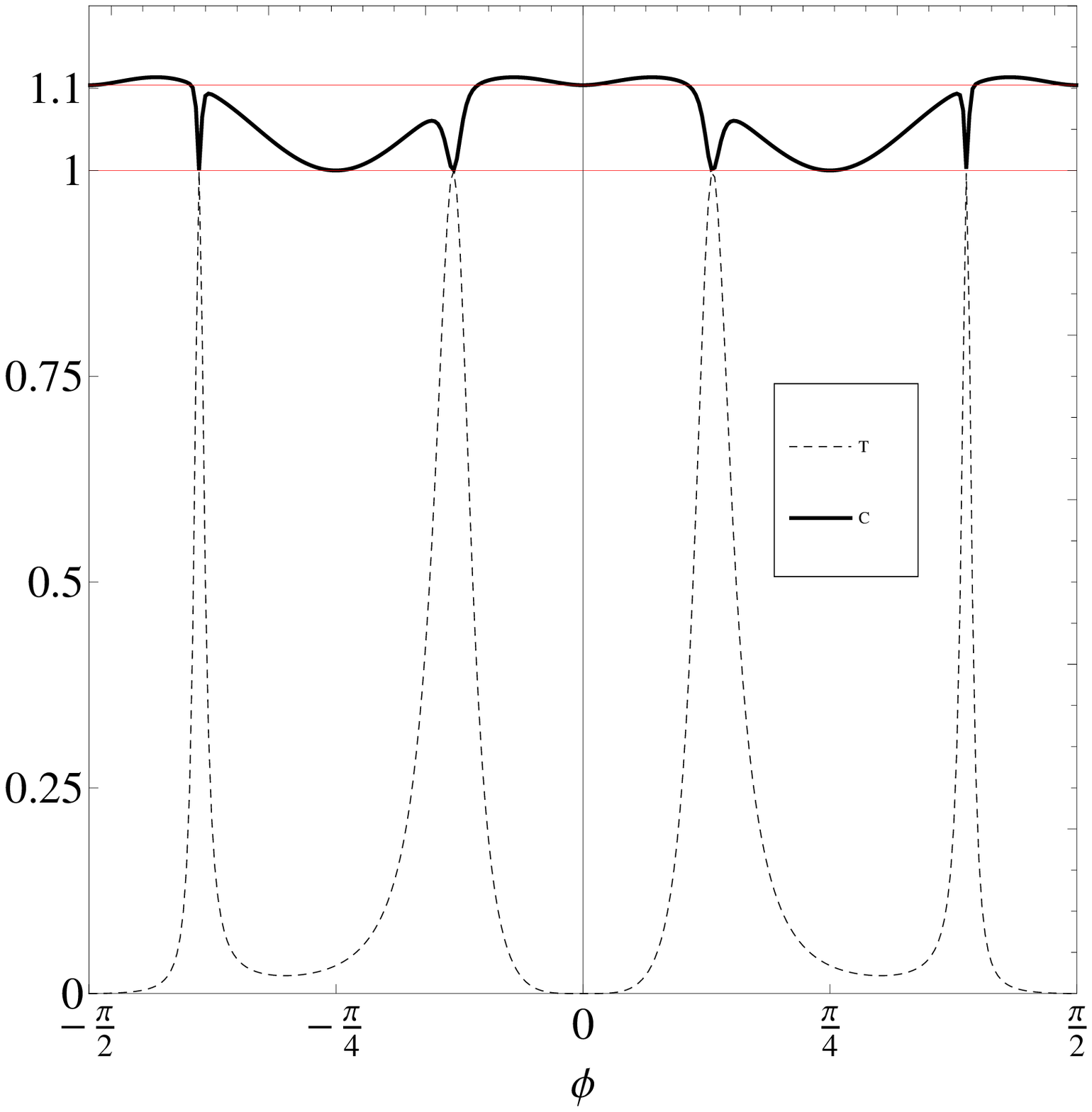}\hskip 5mm\includegraphics[width=7cm]{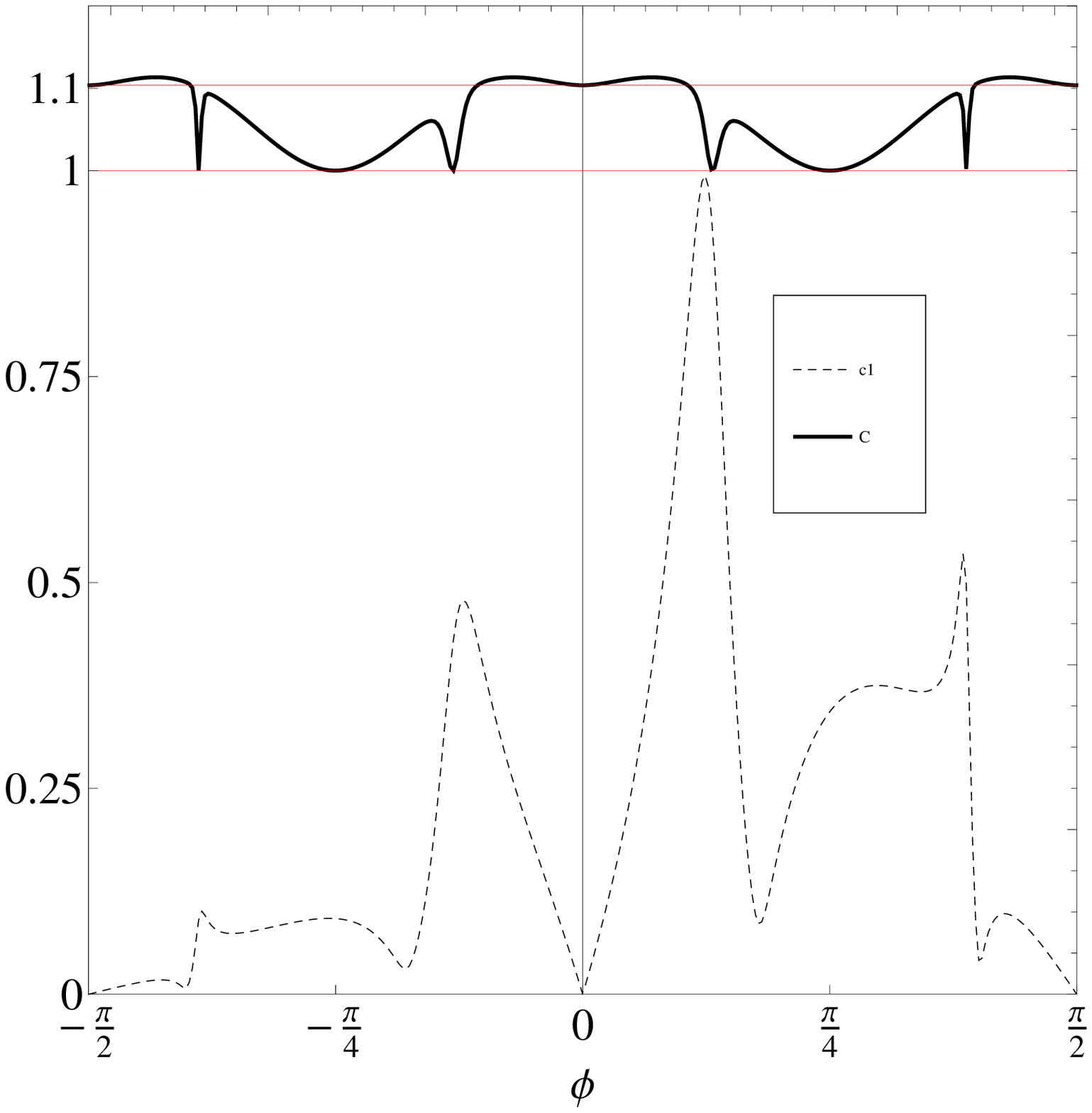}}
\centerline{(a)\hskip 7cm (b)} 
\caption{Statistical complexity $C$, and (a) transmission coefficient $T$ or (b) the coefficient 
$c1=|c_1/(3.1*a_1)|$, in Region I, through a $100$ nm wide barrier vs. the incident angle, $\phi$, for bi-layer graphene. 
The Fermi energy $E$ of the incident electrons is taken $17$ meV. The barrier height $V_0$ 
is $50$ meV. (Levels $C=T=1$ and $C=3/e\simeq 1.1$ are plotted in red).}
\label{fig1}
\end{figure}

In this process, there are no sources or sinks of flux, then the transmission coefficient $T$ is given 
by $T=1-R$. This coefficient $T$ is plotted in Fig. \ref{fig1}a (dashed line) for a given height
$V_0$ of the potential barrier. The anisotropy of the behavior of $T$ is evident in that figure but contrarily 
to the single-layer graphene case, where massless Dirac fermions in normal incidence ($\phi=0$) are perfectly 
transmitted through the barrier \cite{geim2006}, here in the bi-layer graphene case the massive chiral fermions 
are always totally reflected for angles close to $\phi=0$.
The total transparency, $T=1$, can be found for other angles of incidence. This is just the paradoxical 
Klein tunneling in graphene consisting in the penetration of its charge carriers through a high and wide
barrier, $E\ll V_0$ and $2\pi/k_F\ll L$. The calculation of $T$ (and all other magnitudes plotted in the figures) 
has been performed by taking for the electron and hole concentration the typical values used in experiments 
with graphene (see Ref. \cite{geim2006}).

Now, the calculation of the statistical complexity $C$ and the Fisher-Shannon entropy $P$ is presented. 
These magnitudes are the result of a global calculation done on
the probability density $\rho(x,y)$ given by $\rho(x,y)=\Psi^+(x,y)\Psi(x,y)$, taking into account that
the region of integration must be adequate to impose the normalization condition in the two-component 
spinor. As a consequence of having a pure plane wave in the $Y$ axis, the density is a constant in this
direction. Then, without loss of generality, if we take a length of unity in the $Y$ direction, 
the density does not depend on $y$ variable, $\rho(x,y)\equiv \rho(x)$. 
The expressions for densities obtained for the different Regions are cumbersome and are not explicitly given.
To normalize these densities, the length of the integration interval in the $X$ direction 
has been taken to be $[-2\pi/k_x,-\pi/k_x]$, $[0,L]$ and $[L+\pi/k_x,L+3\pi/k_x]$, 
for Regions I, II and III, respectively. Observe that in Regions I and III the integration intervals are taken
separated from the barrier walls at $x=0,L$ in order to diminish in the calculations the effect of the coupling 
between the evanescent and the plane waves.

The statistical complexity $C$ \cite{lopez1995,lopez2002},
the so-called $LMC$ complexity, is defined as
\begin{equation}
C = H\cdot D\;,
\end{equation}
where $H$ is a function of the Shannon entropy of the system and $D$ tells us how far the distribution
is from the equiprobability. Here, $H$ is calculated according to the 
simple exponential Shannon entropy $S$ \cite{lopez2002,shannon1948,dembo1991}, that has the expression, 
\begin{equation}
H = e^{S}\;,
\end{equation}
with
\begin{equation}
S = -\int \rho(x)\;\log \rho(x)\; dx \;.
\label{eq1-S}
\end{equation}
Some kind of distance to the equiprobability distribution is taken for the disequilibrium $D$,
\cite{lopez1995,lopez2002}, that is,
\begin{equation}
D = \int \rho^2(x)\; dx\;.
\label{eq2-D} 
\end{equation}

\begin{figure}[t]
\centerline{\includegraphics[width=7cm]{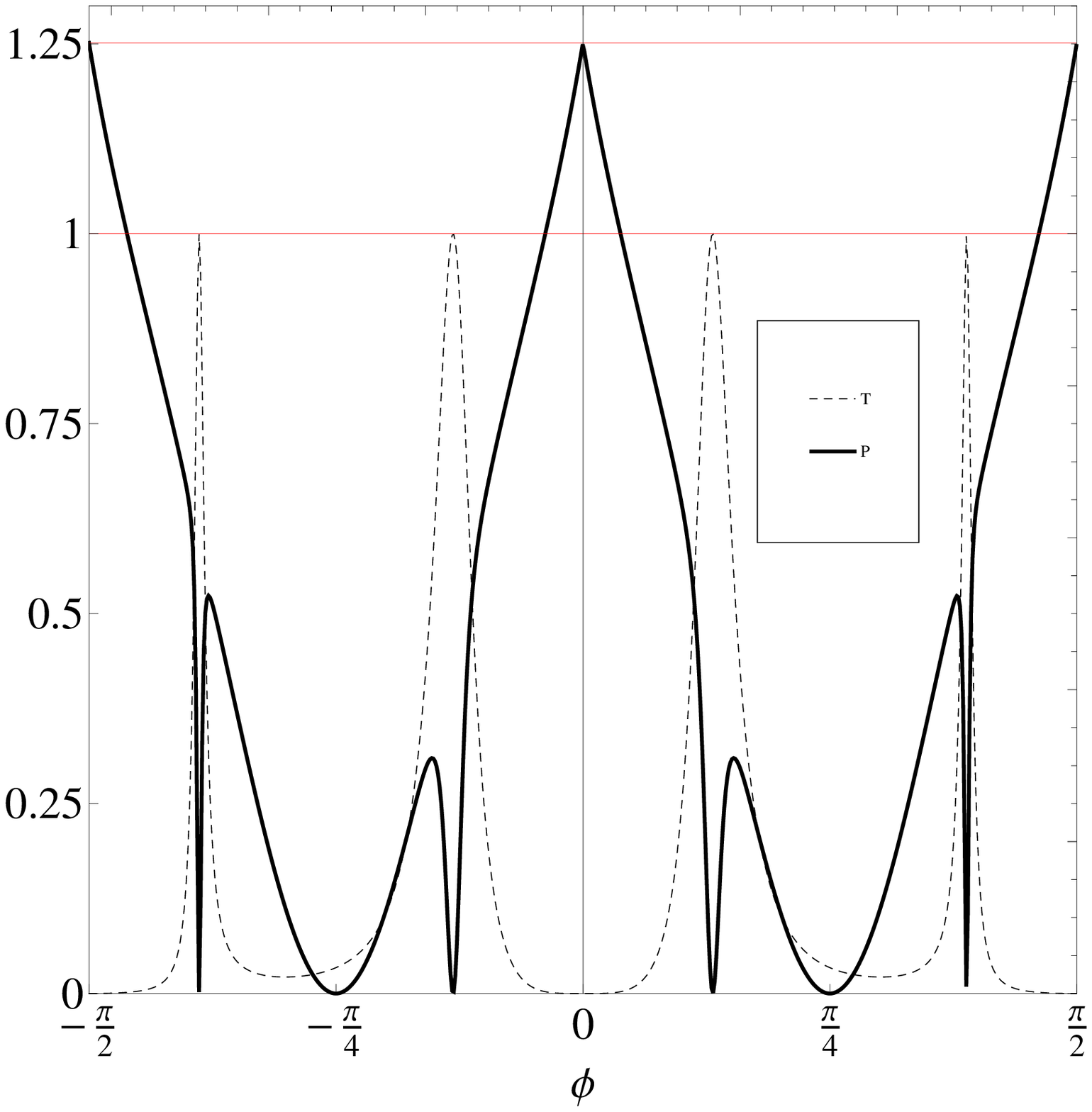}\hskip 5mm\includegraphics[width=7cm]{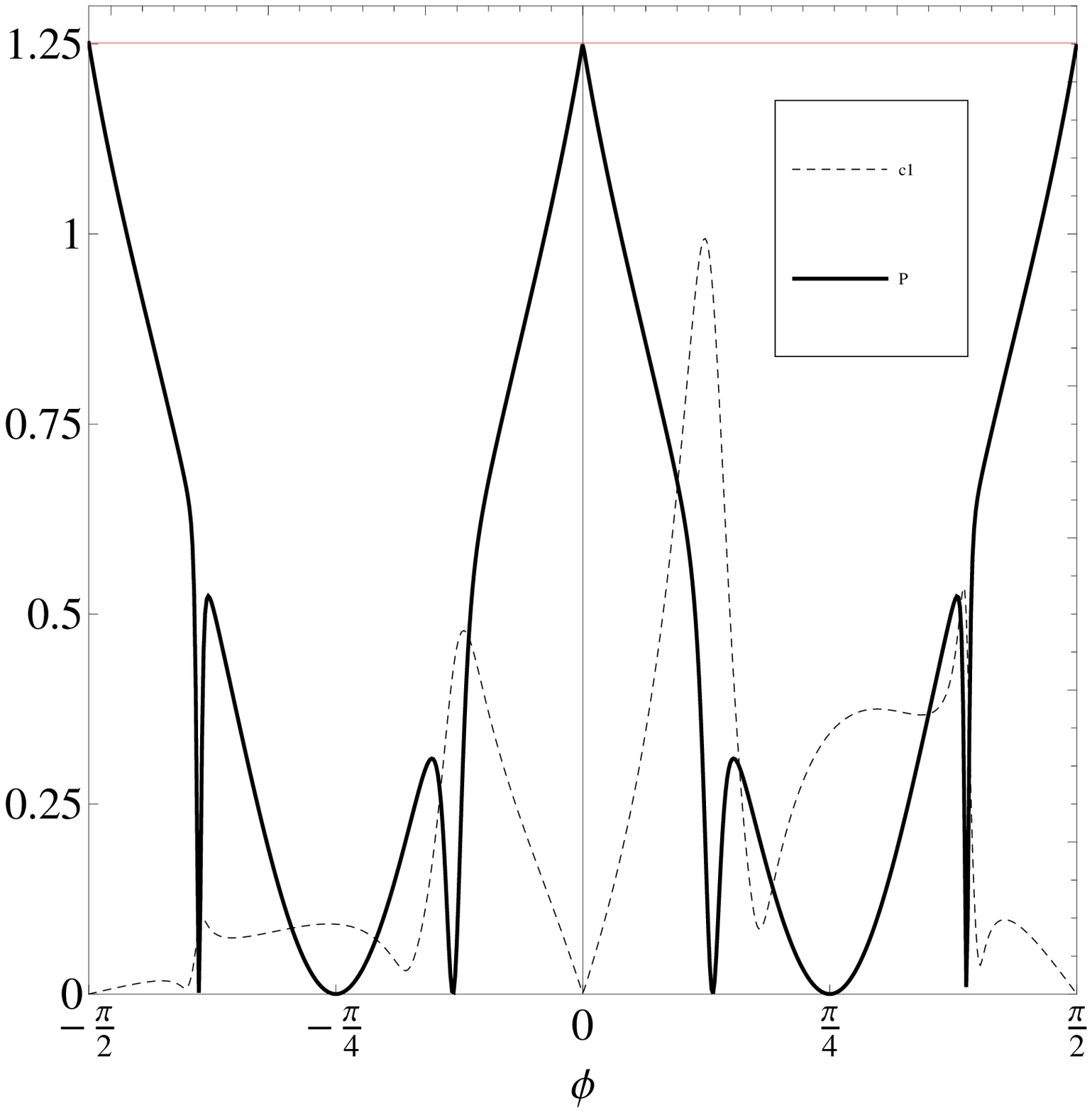}}
\centerline{(a)\hskip 7cm (b)} 
\caption{Fisher-Shannon information $P$, and (a) transmission coefficient $T$ or (b) 
the coefficient $c1=|c_1/(3.1*a_1)|$, in Region I, through a $100$ nm wide barrier vs. 
the incident angle, $\phi$, for bi-layer graphene. The Fermi energy $E$ of the incident 
electrons is taken $17$ meV. The barrier height $V_0$ is $50$ meV.
(Levels $T=1$ and $P=8\pi/e^3\simeq 1.25$ are plotted in red)}
\label{fig2}
\end{figure} 

The Fisher-Shannon information $P$ \cite{romera2004,sen2008} is defined as   
\begin{equation}
P= J\cdot I\,,
\end{equation}
where the first factor is a version of the exponential Shannon entropy \cite{dembo1991}, 
\begin{equation}
J = {1\over 2\pi e}\;e^{2S}\;,
\end{equation}
with the constant $2$ in the exponential selected to have a non-dimensional $P$.
The second factor
\begin{equation}
I = \int {[d\rho(x)/dx]^2\over \rho(x)}\; dx\;,
\end{equation}
is the so-called Fisher information measure \cite{fisher1925}, that quantifies the irregularity
of the probability density.

The statistical complexity $C$ in Region I is plotted in solid line in Fig. \ref{fig1}. 
In contrast to the single-layer graphene transmission behavior where $T=1$ for $\phi=0$, 
in the present case of bi-layer graphene the electrons are totally reflected by the barrier,
$T=0$, under normal incidence.  This gives rise to standing-like waves, due to the interferences 
between the incident and the reflected waves, with complexity $C=3/e\simeq 1.1036$, 
a value that also corresponds to the complexity of the eigenstates of the infinite square well \cite{lopezruiz2009}.
The same behavior is found for parallel incidence $\phi=\pm \pi/2$.
Observe that for incidence angles around $\phi=\pm \pi/4$ the transmission is also near zero,
but in this case, the evanescent waves do not vanish, see the coefficient $c1=|c_1/(3.1*a_1)|$ plotted in Fig. \ref{fig1}b.
The effect of these evanescent waves, $c1\neq 0$, is to destroy the interferences between the incident and reflected waves,
then $C=1$ that corresponds to the complexity of a plane wave. 
For the angles around $\phi=\pm \pi/8$ and  $\phi=\pm \pi/3$, there is total transmission, $T=1$.
It means the electronic flux is just a plane wave, then $C=1$.

In Fig. \ref{fig2}, the behavior of the Fisher-Shannon information $P$ for the Region I 
is plotted in solid line. The transmission coefficient $T$ and the coefficient $c1$ 
of the evanescent waves are also shown in dashed line. For normal and parallel incidence, $\phi=0,\pm \pi/2$, 
standing-like waves are formed around these incidence angle regions, due to interference of the reflected 
and incident waves. The Fisher-Shannon information takes maximum values, $P=8\pi/e^3\simeq 1.2512$, 
in the cases of perfect standing waves, just when a total reflection of the electron flux is achieved. 
In the transparency points, where $T=1$, we have a pure plane wave, it means that the density 
is constant, then the Fisher information is null and so $P$. 
The Fisher-Shannon information also vanishes for incidence angles around $\phi=\pm \pi/4$, 
but in this case the transmission is near zero. This is a consequence of the effect of the evanescent waves 
that destroy the interference pattern between the incident and reflected waves, giving rise to pure 
plane waves, then $P=0$.

\begin{figure}[t]
\centerline{\includegraphics[width=7cm]{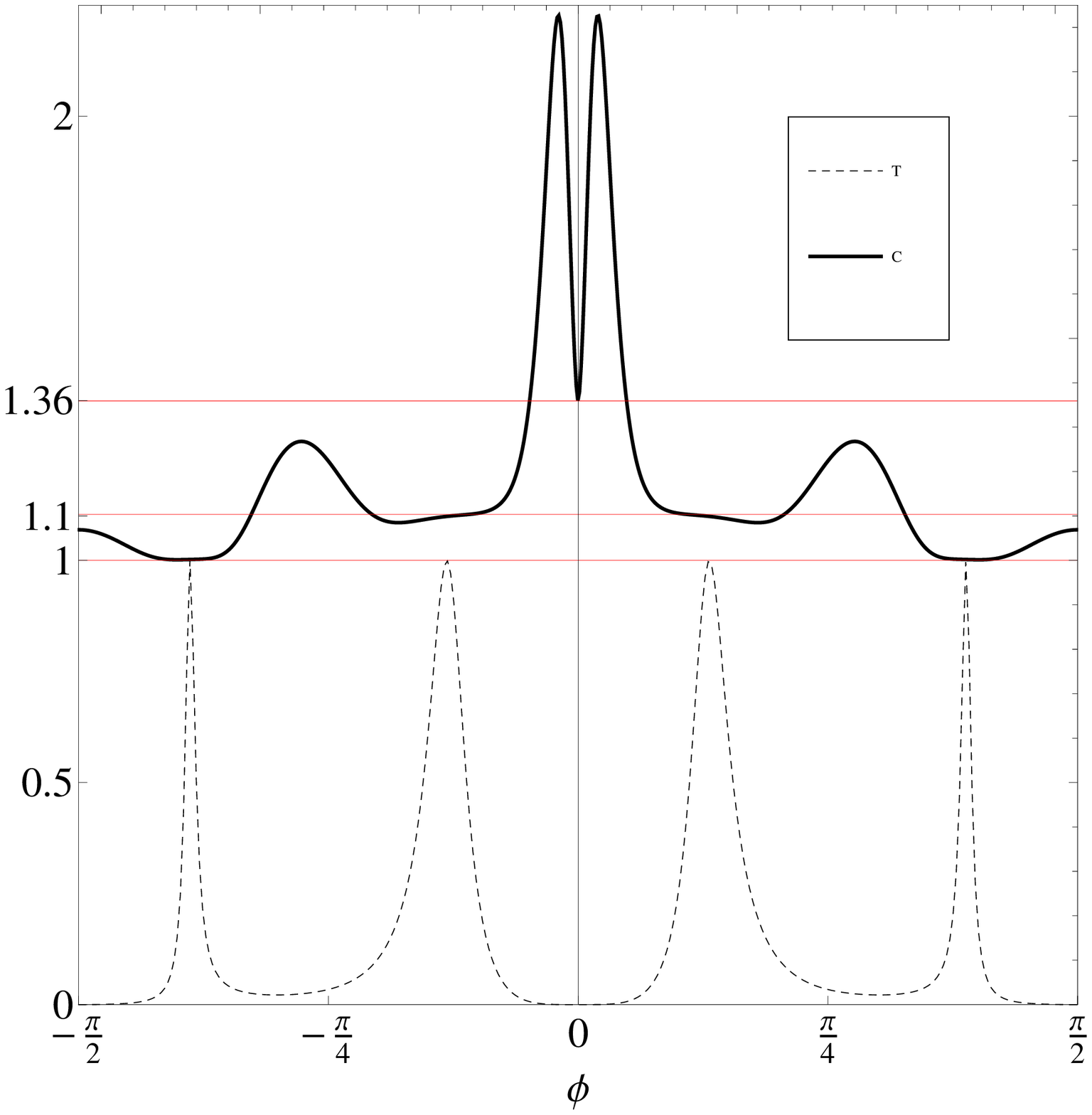}\hskip 5mm\includegraphics[width=7cm]{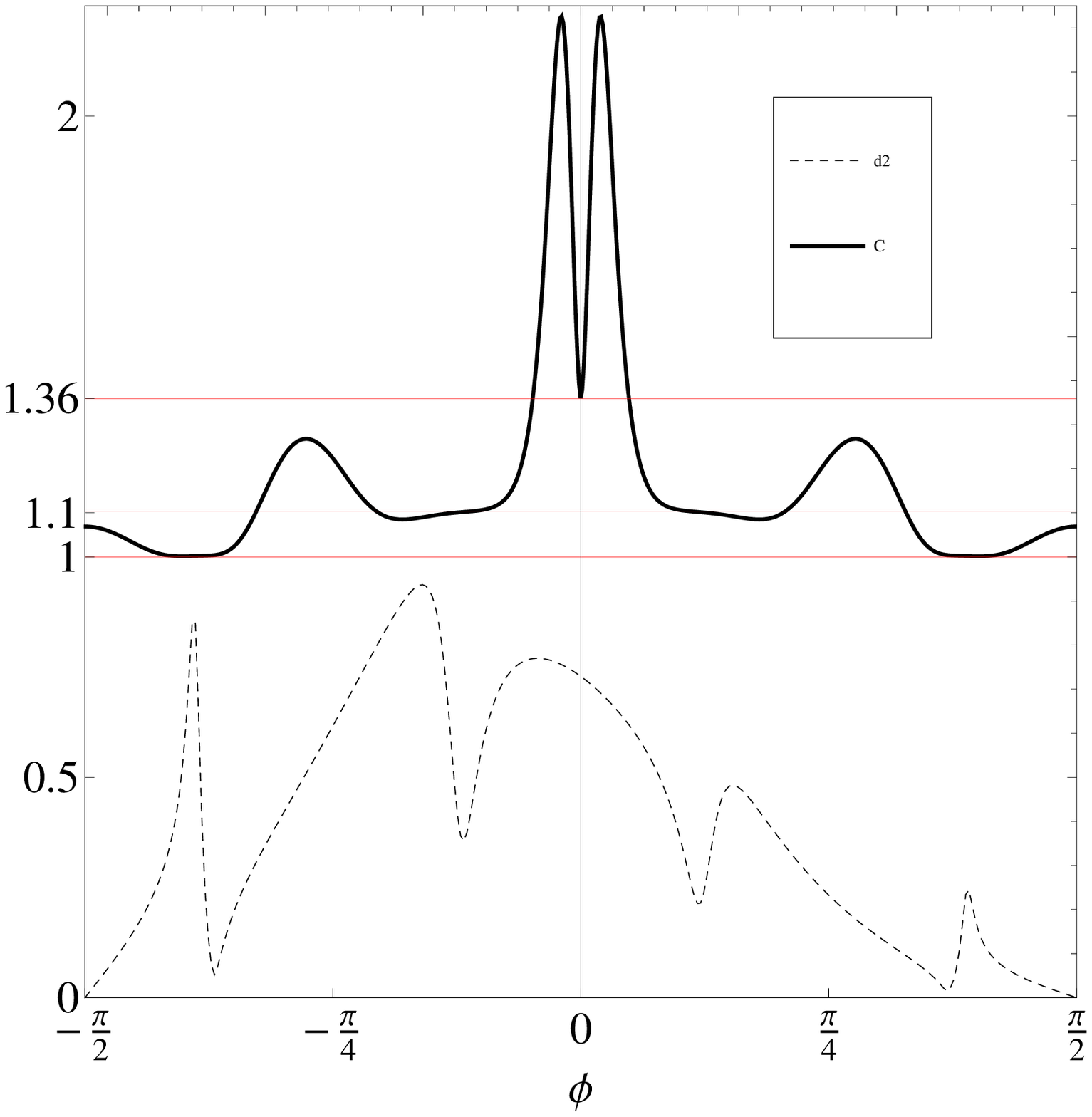}}
\centerline{(a)\hskip 7cm (b)} 
\caption{Statistical complexity $C$, and (a) transmission coefficient $T$ or (b) the coefficient 
$d2=|d_2/(1.6*a_1)|$, in Region II, through a $100$ nm wide barrier vs. the incident angle, 
$\phi$, for bi-layer graphene. The Fermi energy $E$ of the incident electrons is taken 
$17$ meV. The barrier height $V_0$ is $50$ meV. 
(Levels $C=T=1$, $C=3/e\simeq 1.1$ and $C=e/2\simeq 1.36$ are plotted in red).}
\label{fig3}
\end{figure}

\begin{figure}[t]
\centerline{\includegraphics[width=7cm]{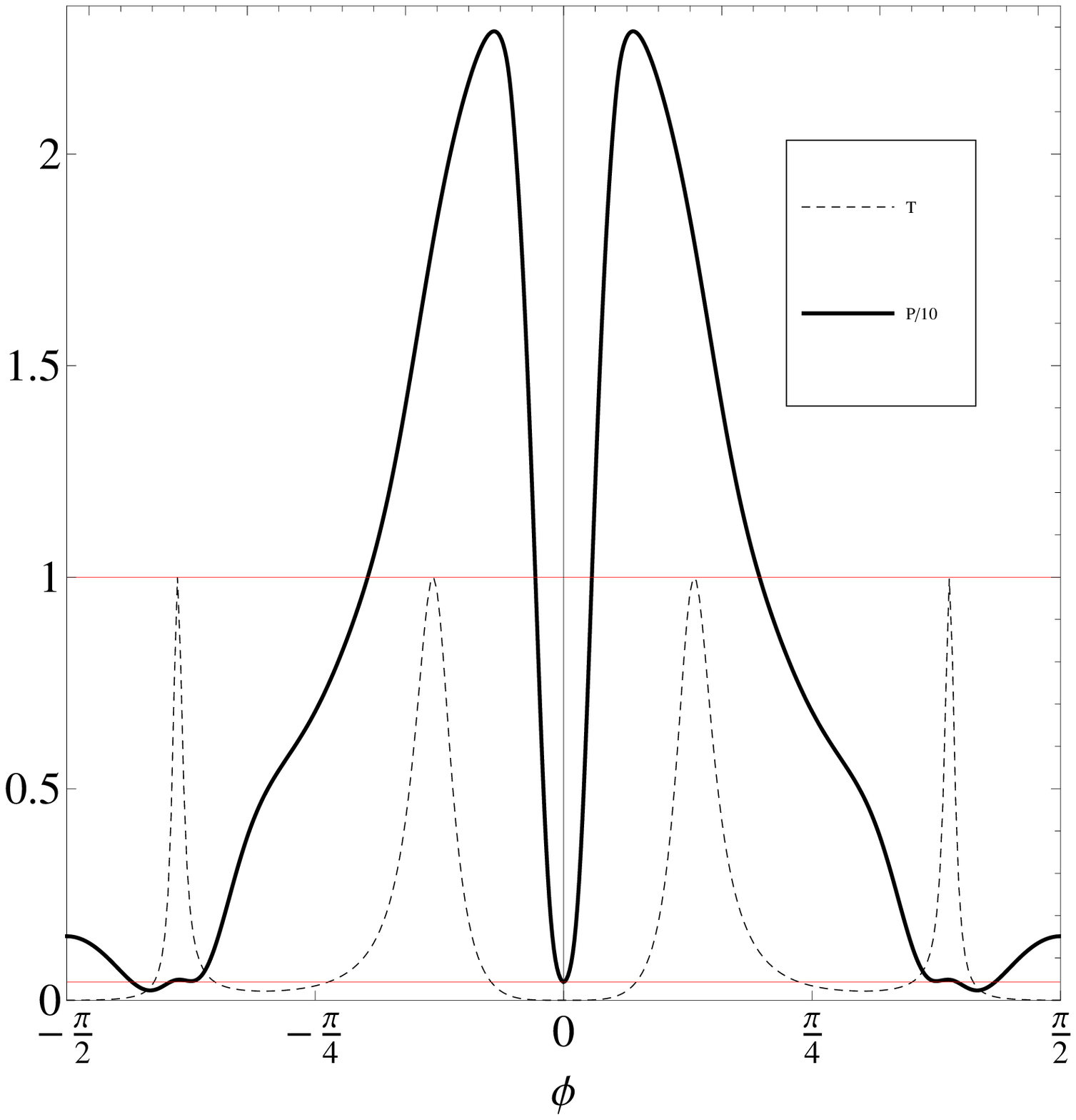}\hskip 5mm\includegraphics[width=7cm]{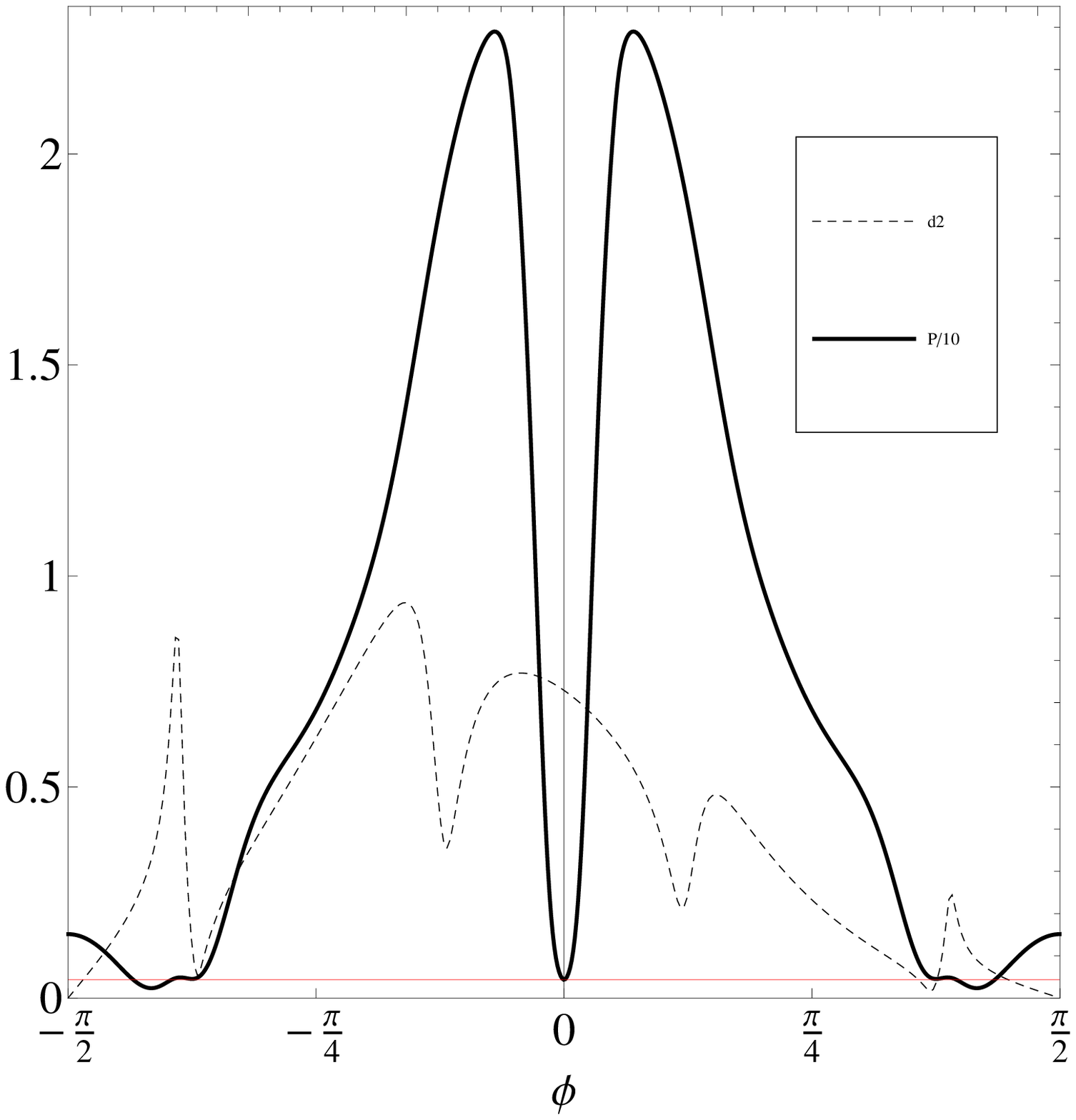}}
\centerline{(a)\hskip 7cm (b)} 
\caption{Fisher-Shannon information $P$, and (a) transmission coefficient $T$ or (b) the coefficient 
$d2=|d_2/(1.6*a_1)|$, in Region II, through a $100$ nm wide barrier vs. the incident angle, $\phi$, 
for bi-layer graphene. The Fermi energy $E$ of the incident electrons is taken $17$ meV. 
The barrier heights $V_0$ is $50$ meV.
(Levels $T=1$ and $P=e/2\pi\simeq 0.43$ are plotted in red).}
\label{fig4}
\end{figure}

The behavior of $C$ for the Region II is plotted in solid line in 
Fig. \ref{fig3}. The transmission coefficient $T$ and the coefficient $d2=|d_2/(1.6*a_1)|$
of the evanescent waves are also shown in dashed line. When the transmission is near zero,
for $\phi=0$ and $\phi$ around $\pm \pi/4$,
the wave function in the interior of the barrier is essentially the evanescent wave.
This means that the complexity of a decreasing exponential wave is approached, i.e. $C=e/2\simeq 1.36$.
In the transparency points, $C$ takes the value of a standing wave $C\simeq 1.1036$
or the value $C=1$  that corresponds to a plane wave, depending on the strength of the evanescent 
waves, if they are stronger enough to allow the formation of standing waves or not, respectively. 
See the plot of the coefficient $d2$ in Fig. \ref{fig3}b.
This last behavior is recovered in the plot of $P$ in Fig. \ref{fig4},
where the quantity $P/10$ is displayed.
$P$ takes its maximum near the first transparency point where oscillations are present
in the standing waves and $P$ takes its minimum near the second transparency point where
a constant density is obtained from the planes waves present inside the barrier.
For $\phi=0$, the Fisher-Shannon information taken by the evanescent wave, 
just an exponential distribution, is $P=e/2\pi\simeq 0.43$.

\begin{figure}[t]
\centerline{\includegraphics[width=7cm]{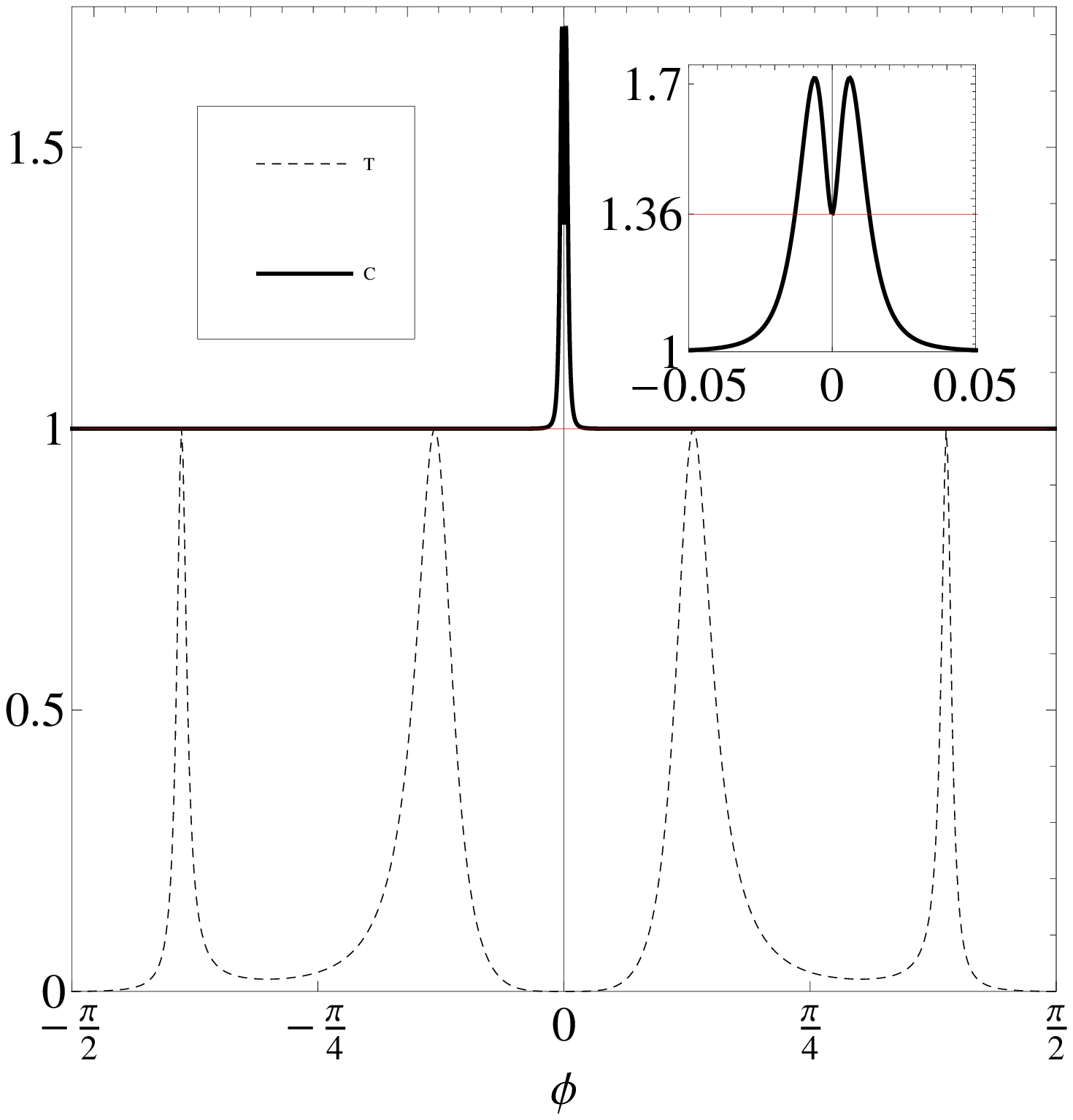}\hskip 5mm\includegraphics[width=7cm]{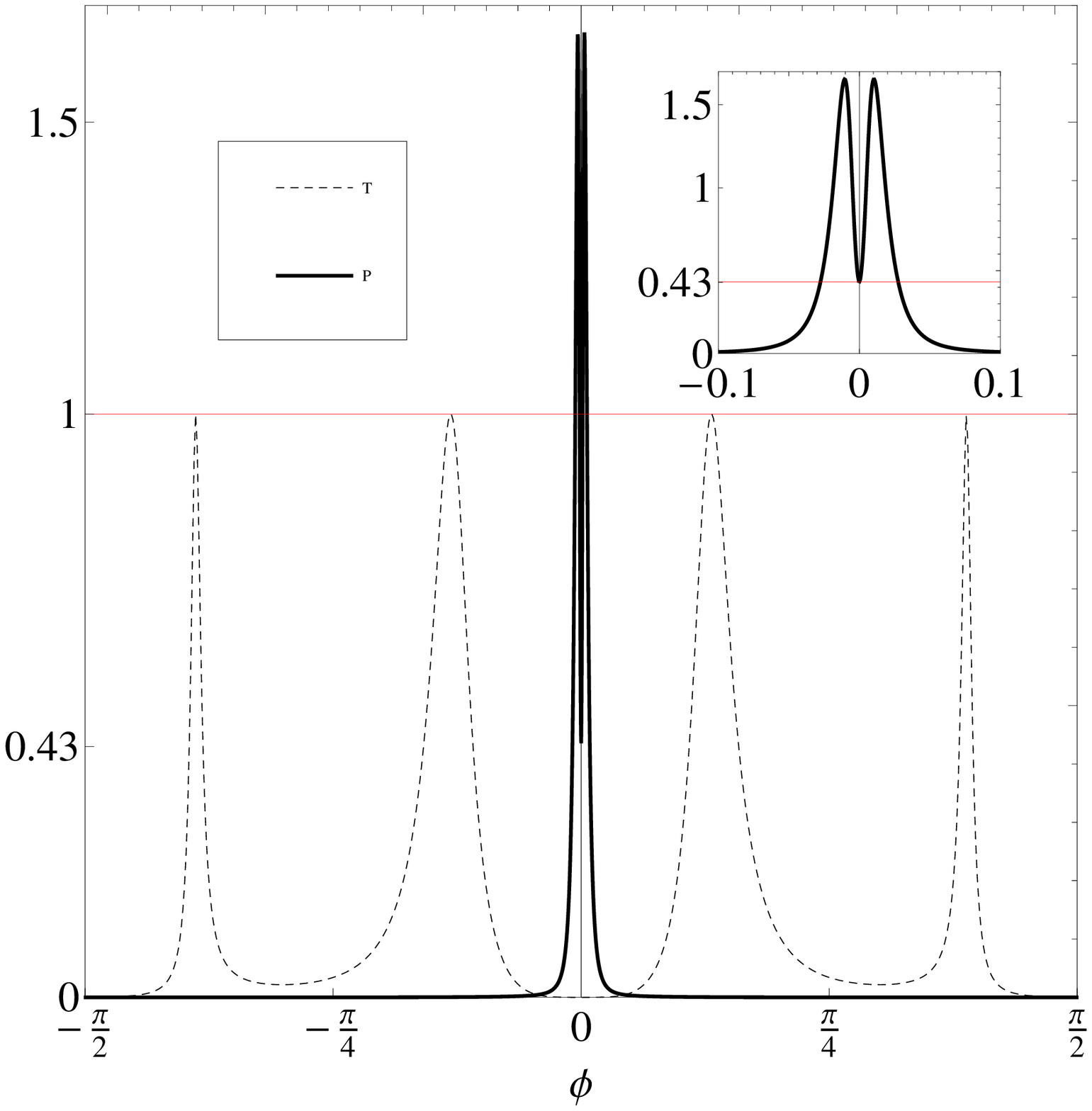}}
\centerline{(a)\hskip 7cm (b)} 
\caption{Transmission coefficient $T$, and (a) statistical complexity $C$ or (b) Fisher-Shannon information $P$, 
in Region III, through a $100$ nm wide barrier vs. the incident angle, $\phi$, for bi-layer graphene. 
The Fermi energy $E$ of the incident electrons is taken $17$ meV. The barrier heights $V_0$ is $50$ meV.
Insets are zoom of (a) C and (b) P around $\phi=0$. (Levels $C=T=1$, $C=e/2\simeq 1.36$ and 
$P=e/2\pi\simeq 0.43$ are plotted in red).}
\label{fig5}
\end{figure}

In the Region III, there is not reflected wave then no interference capable to create
standing waves is possible. Only the evanescent and the right traveling waves are present. 
In the limit of perfect reflection for $\phi=0$, 
there is not transmission of energy through the barrier and
only the evanescent wave is found. Then, $C=e/2\simeq 1.36$ and $P=e/2\pi\simeq 0.43$ for $\phi=0$.
Far from the tiny angle region around $\phi=0$, the traveling wave is the main contributor in the 
density, it means that the density is near a constant for any angle of incidence, 
and $C$ and $P$ take their minimum 
values, $C=1$ and $P=0$, as it can be seen in Fig. \ref{fig5}.

In conclusion, following the steps given in a previous work \cite{sanudo2014} on single-layer graphene, 
we have performed the calculation of the statistical complexity 
and the Fisher-Shannon information in a problem of electron scattering 
on a potential barrier in bi-layer graphene. In this case, the evanescent waves 
are an important ingredient of the wave function near the limits of the barrier.
The influence of these decreasing exponential waves in the statistical indicators
has been computed. Concretely, it has been observed that even when the transmission
through the barrier is near zero, these statistical magnitudes can take their minimum values,
a situation that was not present in the single-layer graphene, where the minimum values 
for $C$ and $P$ were taken only on the angles with complete transmission.  
This put in evidence that these entropic measures can also be useful to discern
certain physical differences in similar quantum problems, as for instance the case of
the Klein tunneling in single and bi-layer graphene.

\section*{Acknowledgements}
 J.S. thanks to the Consejer\'ia de 
 Econom\'ia, Comercio e Innovaci\'on of the Junta de Extremadura (Spain) for 
 financial support, Project Ref. GRU09011.

\end{document}